\documentclass[aps,prb,reprint,showpacs,superscriptaddress,longbibliography]{revtex4-1}
\usepackage{mathtools,amssymb,graphicx,units}
\usepackage[usenames,dvipsnames]{color}
\usepackage[plainpages=false,pdfpagelabels,colorlinks=true,linkcolor=red,urlcolor=blue,citecolor=blue,pdftitle={Title},pdfauthor={},pdfdisplaydoctitle=true,pdfduplex=DuplexFlipLongEdge]{hyperref}

\graphicspath{{./}}

\begin{document}

\title{Dynamical localization and the effects of aperiodicity in Floquet systems}

\author{Tilen \v Cade\v z}
\email{tilen.cadez@csrc.ac.cn}
\affiliation{Beijing Computational Science Research Center, Beijing, 100193, China}
\author{Rubem Mondaini}
\email{rmondaini@csrc.ac.cn}
\affiliation{Beijing Computational Science Research Center, Beijing, 100193, China}
\author{Pedro D. Sacramento}
\email{pdss@cefema.tecnico.ulisboa.pt}
\affiliation{CeFEMA, Instituto Superior T\'{e}cnico, Universidade de Lisboa, Av. Rovisco Pais, 1049-001 Lisboa, Portugal}
\affiliation{Beijing Computational Science Research Center, Beijing, 100193, China}

\begin{abstract}
We study the localization aspects of a kicked non-interacting one-dimensional (1D) quantum system subject to either time-periodic or non-periodic pulses. These are reflected as sudden changes of the on-site energies in the lattice with different modulations in real space. When the modulation of the kick is incommensurate with the lattice spacing, and the kicks are periodic, a well known dynamical localization  in real space is recovered for large kick amplitudes and frequencies. We explore the universality class of this transition and also test the robustness of localization under deviations from the perfect periodic case. We show that delocalization ultimately sets in and a diffusive spreading of an initial wave packet is obtained when the aperiodicity of the driving is introduced.
\end{abstract}


\pacs{72.15.Rn, 05.45.Mt, 64.70.Tg}

\maketitle

\section{Introduction}
Quantum localization is a topic of long-standing interest that is manifest on a broad range of contexts, either in time-independent systems~\cite{Evers_Mirlin_2008} as well as in time-periodic ones~\cite{Eckardt_2017}. The possibility of experimentally testing the theoretical predictions with highly tunable experiments, as provided by optical lattice setups~\cite{Billy_2008, Roati_2008} or in photonic lattices~\cite{Schwartz_2007, Lahini_2008}, has led to a boost on the investigation of this phenomenon. Non-interacting quantum systems with static quenched disorder~\cite{Anderson_58} (or quasi-disorder~\cite{Aubry_80}) are well known to display localization of all its single-particle states in low dimensions~\cite{Abrahams1979}. In the realm of time-periodic driven quantum systems, frequently called Floquet quantum systems, dynamical localization, where in spite of the periodic pumping of energy into the system its total energy remains maximally bounded, has been also widely observed, as, e.g., in two-level systems~\cite{Grossmann_1991} and in quantum kicked rotors~\cite{Casati_79, Haake, Fishman_1982, Moore_1995, Chabe_2008}. As another remarkable example of the theory and experimental connection in the investigation of dynamical localization is the case of a charged particle in a lattice subjected to a sinusoidal force in time~\cite{Dunlap_1986}, that was later realized with ultracold Bose-Einstein condensates~\cite{Lignier_2007, Eckardt_2009} trapped in optical lattices.

In ergodic interacting systems, on the other hand, it has been shown that a periodic driving leads the system to a featureless \textit{infinite temperature} regime, characterized by a non-equilibrium steady-state that is locally identical to a state with maximal Gibbs entropy~\cite{Luitz_2017}. Thus, it typically leads to thermalization (and therefore absence of localization) once the periods of the driving are sufficiently large~\cite{DAlessio_2013, DAlessio_2014, Lazarides_2014, Regnault_2016}. Other integrable systems, which possess an extensive number of integrals of motion, do not heat up and exhibit localization in energy space at infinite times~\cite{Lazarides_2014b,Gritsev_2017} even in the case of specific types of aperiodic driving~\cite{Nandy17}. A third class of interacting systems under periodic driving are the ones with an emerging integrability as in the case of disordered systems displaying the many-body localization phenomenon \cite{Lazarides_2015, Ponte_2015, Abanin_2016}, that were recently experimentally investigated~\cite{Bordia_2017}. In these cases, the (many-body) localization under the driving is stable only at large frequencies and a fully mixed featureless state entails after increasing the drive period.

Here, our purpose is to study the dynamical localization (and its breakdown) of non-interacting lattice systems which are driven by global pulses affecting the site energies. These global pulses are instantaneous and hereafter we refer to them as kicks in the lattice. This has been the recent focus of studies when the kicks in real space are either incommensurate with the lattice periodicity~\cite{Qin_14}, or when they are commensurate but with wavelengths which are twice the lattice spacing~\cite{Agarwala_17}. In both cases, it leads to dynamical localization, characterized by the halt of the spreading of an initial wave packet when the frequency of the kicks is large and their magnitudes are either sufficiently large~\cite{Qin_14} or when they meet resonant conditions~\cite{Agarwala_17}. Our goal is to provide a broader picture of the dynamical localization in real space with a systematic study of two different kick protocols in a tight-binding Hamiltonian, observing its manifestation. Besides this, we aim to investigate whether the dynamical localization is robust when the periodic driving is slightly altered, say by performing drivings at non-fixed periods. 

This latter aspect has been experimentally investigated in the cases of the atomic~\cite{Oskay_2003} and molecular quantum kicked rotors~\cite{Bitter_2016, Bitter_2017}, whose dynamical localizations are obtained in momentum and angular momentum, respectively. Deviations on the periods of the driving were shown to lead to decoherence, ultimately recovering ergodic behavior. We test these ideas in a lattice model where the localization is, however, manifested in real space.

The presentation is structured as follows: In Sec.~\ref{sec:model_methods} we introduce the model and describe the basics of Floquet systems. Section~\ref{sec:periodic} explores the dynamical localization under periodic driving and the universality class of the transition, while Sec.~\ref{sec:aperiodic} generalizes the problem for the situations where the kicks in the lattice are no longer equally spaced in time. Lastly, Sec.~\ref{sec:summary} summarizes our findings.

\section{Model and methods}
\label{sec:model_methods}

We consider a 1D model of spinless fermions, in a lattice with periodic boundary conditions, whose Hamiltonian reads
\begin{eqnarray}
\hat{H} = \hat{H}_0 + \lambda {\sum_{\tau} \delta(t-T_{\tau})} \, \hat{V},
\label{eq:hamilt}
\end{eqnarray}
where $\hat{H}_0 = -J \sum_{i=1}^L \, \hat{c}_i^{\dagger} \hat{c}_{i+1} + \hat{c}_{i+1}^{\dagger} \hat{c}_{i}$ is the kinetic energy, with $\hat{c}_i^{\dagger}$ ($\hat{c}_i$) a particle creation (annihilation) operator at site $i$ and $J$, the nearest-neighbor hopping amplitude, that we set to unity ($J = 1$), defines the energy scale of the problem. The second term, $\hat{V} = \sum_{i=1}^L \, V_i \, \hat{c}_i^{\dagger} \hat{c}_{i}$, is the potential which is applied onto the system at times $T_{\tau}$. These act as kicks in time by quenching the on-site energies of the lattice whose maximal amplitude is given by $\lambda$. Note that $\tau$ is an integer, counting the number of applied kicks.

We study two types of kicks where the site energies $V_i$ are either uncorrelated, as provided by an Anderson-like potential, $V_i^{\rm A}$, with randomly selected energies within a symmetric interval $[-1,1]$, or in correlated cases as in a quasi-periodic disordered case, $V_i^{\rm AA}= \cos(2 \pi \alpha i  + \varphi)$, which emulates the Aubry-Andr\'e potential~\cite{Harper_55, Aubry_80, Artuso94}, acting only at instants of time $T_{\tau}$. We set $\alpha$ as the inverse golden ratio $(\sqrt{5}-1)/2$ and the phase $\varphi\in[0,2\pi)$ allows a ``disorder'' average which can reduce statistical and finite-size effects.

We will start by describing the physics that results on the application of these kicks periodically in time, i.e., when $T_{\tau} = \tau T$. In this case, the Hamiltonian~(\ref{eq:hamilt}) is periodic and the Floquet formalism will aid in the stroboscopic description of, e.g., the time evolution of a wave packet to determine whether dynamical localization takes place or not.

\subsection*{Floquet basics}

According to the Floquet theorem, the time-evolution operator of a periodically driven system, described by a time-periodic Hamiltonian $\hat{H}(t+T) = \hat{H}(t)$ with period $T$, can be written (in units where $\hbar=1$) as~\cite{Shirley_65, Sambe_73, Stockmann, Grifoni_98, Bukov_15} 
\begin{eqnarray}
\hat{U}(t) = \hat{P}(t) \, e^{- {\rm i} \hat{H}_{\mathrm{eff}} t},
\end{eqnarray}
where $\hat{P}(t) = \hat{P}(t+T)$ is a time-periodic operator with $\hat{P}(0) = \hat I$ being the identity and $\hat{H}_{\mathrm{eff}}$ is the time-independent Floquet Hamiltonian. After one driving period, the time-evolution operator is 
\begin{eqnarray}
\hat{U}(T) = e^{- {\rm i} \hat{H}_{\mathrm{eff}} T} = \sum_m e^{- {\rm i} \varepsilon_m T} \, | \theta^m \rangle \langle \theta^m |,
\label{eq:evol_op}
\end{eqnarray}
where $\varepsilon_m$ are the Floquet quasienergies, which are connected to the eigenvalues of $\hat{U}(T)$, and $| \theta^m \rangle$ are its eigenvectors, that are stationary states at stroboscopic times of the driven system. Note, however, the ambiguity in the definition of the effective Hamiltonian in (\ref{eq:evol_op}) since the Floquet quasi-energies can be shifted by a multiple of $\omega = 2 \pi/T$ without affecting the eigenvalues of $\hat{U}(T)$. In the case of time-periodic kicks, the time-evolution operator can be written as a quantum map~\cite{Berry_79}, $\hat{U}(T) = e^{-{\rm i} \hat{H}_0 T} e^{-{\rm i} \lambda \hat{V}}$. In general, there is no guarantee that one can write down a closed form of the effective time-independent Hamiltonian~\cite{Bukov_15} which could potentially represent an undriven physical system. For that, the Floquet Hamiltonian should be in a form of a local operator and this issue is related to the convergence of the Magnus expansion often used to explicitly obtain $\hat{H}_{\rm eff}$ in the high-frequency limit ($1/T\gg1$). Here, on the other hand, the simple form of the stroboscopic evolution operator allows us to write down the Floquet Hamiltonian by making use of the Baker-Campbell-Hausdorff (BCH) formula, $\exp{\hat X}\exp{\hat Y} = \exp\{\hat X + \hat Y + \frac{1}{2}[\hat X,\hat Y] + \frac{1}{12}[[\hat X,\hat Y],\hat Y]$ $+ \frac{1}{12}[\hat X,[\hat X,\hat Y]] + \cdots\}$, as
\begin{eqnarray}
 \hat H_{\rm eff} &=& \hat H_0 + \frac{\lambda}{T}\hat V - {\rm i} \frac{\lambda}{2}[\hat H_0, \hat V] \nonumber \\
 &-& \frac{\lambda^2}{12}[[\hat H_0,\hat V],\hat V] - \frac{T\lambda}{12}[\hat H_0,[\hat H_0,\hat V]] + \cdots.
 \label{eq:H_eff}
\end{eqnarray}
In the high-frequency $(1/T\gg1)$ and small kick-amplitudes $(\lambda \ll 1)$ limits, the effective Hamiltonian thus assumes a closed form,
\begin{equation}
 \hat H_{\rm eff} = \hat H_0 + \frac{\lambda}{T} \hat V.
 \label{eq:floquet_hamil}
\end{equation}
This effective time-independent Hamiltonian is reminiscent of well known models of static quenched disorder as in the Anderson~\cite{Anderson_58} and Aubry-Andr\'e~\cite{Aubry_80} models, in the situation where the associated kick-operator $\hat V$ is now static and specified by $V_i^{\rm A}$ and $V_i^{\rm AA}$, respectively. Note, however, that the disorder amplitudes become renormalized by the driving period $T$. While the BCH formula gives us insights on the  problem in the high-frequency and small kick-amplitude limits, for general parameter values, we use exact diagonalization of the time-evolution operator to probe localization for different parameters and in regimes not necessarily obeying these limits.

\begin{figure}[!t]
\includegraphics[width=0.95\columnwidth]{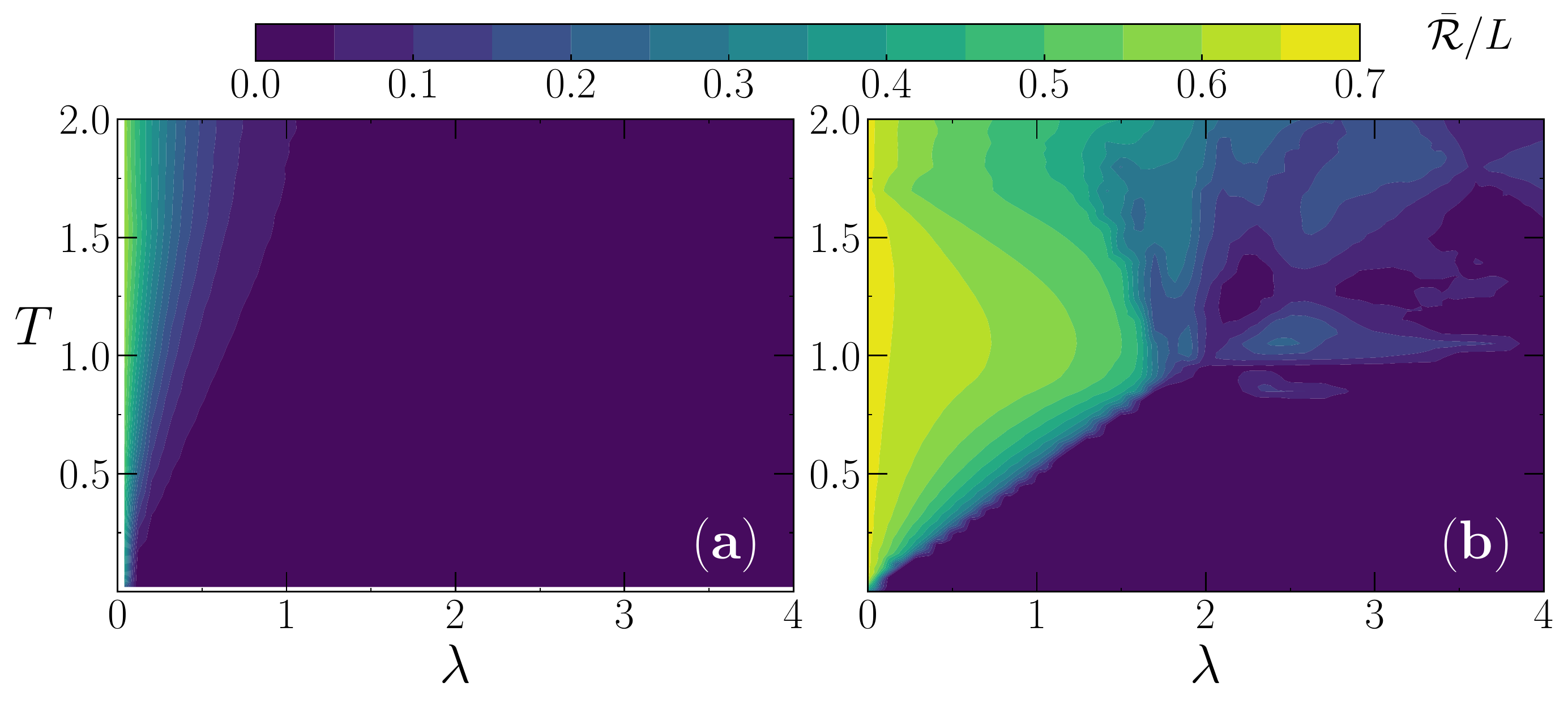}
\caption{(Color online) The mean IPR $\overline{\cal R}/L$ as a function of the potential strength $\lambda$ and period $T$ for Anderson (a) and Aubry-Andr\'e (b) periodic potential kicks as a contour plot. The purple (dark) regions denote the regime where non-ergodicity, and hence dynamical localization, takes place while the yellow (light) regions where the system is ergodic and delocalized. The system size is $L = 1000$ and we average over 10 realizations of disorder.}
\label{fig:1}
\end{figure}

\section{Dynamical localization in periodically kicked systems}
\label{sec:periodic}
\subsection{Non-ergodicity in the eigenstates}
Dynamical localization can be probed in different ways. From the experimental point of view, for example, one is interested in checking if an initially localized wave packet spreads or not in real space after subsequent kicks in the system. While this can also be investigated theoretically~\cite{Qin_14}, one can easily  infer localization (delocalization) by just recalling its connections with non-ergodic (ergodic) properties~\cite{Haake}. In turn, the level of ergodicity can be quantified by the inverse participation ratio (IPR) of the eigenvectors of the time-evolution operator after one period (the Floquet operator) $\hat{U}(T)$, which we define as ${\cal R}_m = 1/\sum_i | \theta^{m}_i |^4$. In Fig.~\ref{fig:1}, we report the average IPR, $\overline{\cal R} = \langle 1/\sum_i | \theta^{m}_i |^4 \rangle_{m,r}$, as a function of $\lambda$ and $T$ for the two kick protocols considered. Here, $\langle \cdot \rangle_{m,r}$ denotes the average over all the eigenstates $| \theta^{m} \rangle = \sum_i \theta^{m}_i | i \rangle$ in the site basis $| i \rangle$, also averaged over different realizations $r$, where different realization consist of different choices of random onsite energies and different $\varphi$'s, for Anderson and Aubry-Andr\'e potentials, respectively. 

The average IPR quantifies the average spreading of the eigenvectors in real space. In the absence of the potential ($\lambda = 0$), all $| \theta^{m} \rangle$ are plane waves, for which $\overline{\cal R} = L$ and for general delocalized states $\overline{\cal R}\sim {\cal O}(L)$, whereas for perfectly localized states $\overline{\cal R} = 1$. It has the advantage of not relying on any specific details of initial states in which the system is prepared but it is, however, basis-dependent. That does not constitute a problem since we aim in investigating dynamical (de)localization in the real-space (site) basis which is also the computational basis.

In Figs.~\ref{fig:1}(a) and ~\ref{fig:1}(b), where we report the results for the two disordered kick-protocols ($V_i^{\rm A}$ and $V_i^{\rm AA}$, respectively), we notice that localization, $\overline{\cal R} \sim {\cal O}(1)$, is robust in the high-frequency regime. In fact, with the prescription of the effective Floquet Hamiltonian (Eq.~\ref{eq:floquet_hamil}) valid when $T\ll 1$ and $\lambda \ll 1$, we see that the critical values for the onset of dynamical localization are $\lambda/T = 0^+$ and $\lambda/T = 2$, that correspond to the onset of localization in the static disordered Anderson and Aubry-Andr\'e models in one-dimension, respectively, when the disorder energy scale is renormalized by $T$. When the frequency of the kicks decreases, the effective Hamiltonian description given by (\ref{eq:floquet_hamil}) no longer holds and a more complicated outcome of the ergodicity ensues with higher order terms in (\ref{eq:H_eff}) being necessary to explain its details.

\begin{figure}[!t]
\centering
\includegraphics[width=0.999\columnwidth]{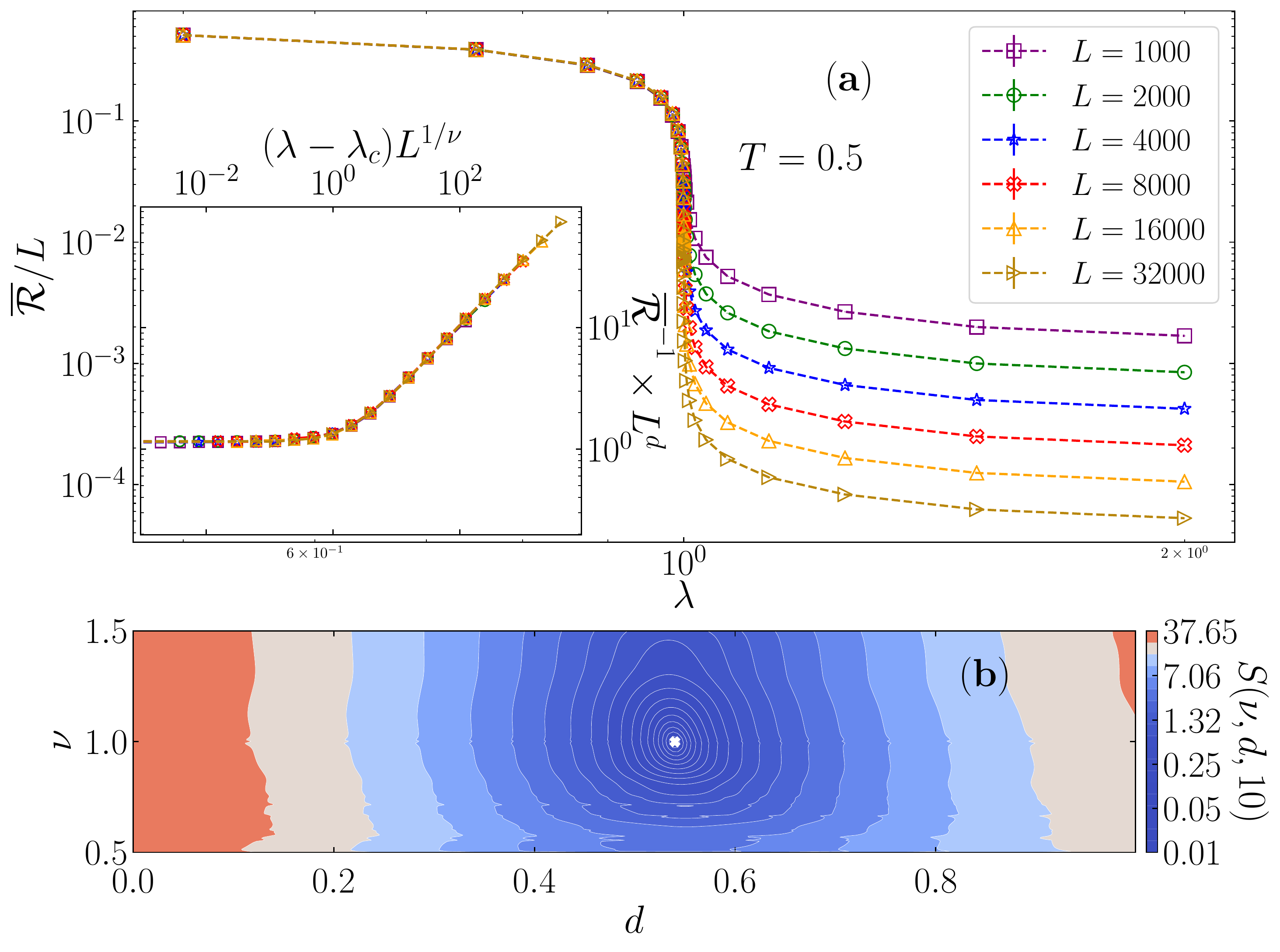}
\caption{(Color online) (a) The average IPR $\overline{\cal R}/L$ for periodically kicked AA model at fixed period $T = 0.5$ for different system sizes up to $L = 32000$. Inset shows the data collapse of the finite-size scaling. (b) Contour plot of the fitting error as a function of fractional dimensionality $d$ and the critical exponent $\nu$, which gives $d = 0.54\pm0.01$ and $\nu = 1.00\pm0.01$. The polynomial fitting used is of the order $n = 10$.}
\label{fig:2}
\end{figure}

To further understand the connection between the Floquet Hamiltonian and the static disordered correspondents, we explore the critical exponents of the dynamical localization at the transition point, $\lambda_c/T_c = 2$, for the case of the Aubry-Andr\'e potential. Thus, we study the scaling properties of the average IPR and compute the critical exponent $\nu$, which describes the divergence of the localization length $\xi$ as $\xi \sim \left|\lambda-\lambda_c\right|^{-\nu}$ in the vicinity of the transition. There, the average IPR exhibits a scaling invariant form as $\overline{\cal R}^{-1} L^{d} \sim f\left[(\lambda-\lambda_c)L^{1/\nu}\right]$, where $d$ is related to the multifractal dimension of the wave function~\cite{Hiramoto_92,Li_Pixley_2016}. In Fig.~\ref{fig:2}(a), we show how the IPR is abruptly reduced (at $\lambda_c = 2T_c$) after increasing the kick strength for different system sizes. The collapsed scaling form is depicted in the inset. To simultaneously extract the critical exponents $\nu$ and $d$ that resulted in this plot, we systematically use the error stemming from a high-order polynomial fitting of the points correspondent to different system sizes, for a range of values of $\nu$ and $d$. When the points coincide in a smooth curve, for the best set of parameters $\{\nu,d\}$, the associated error is small. This linear-square fitting error $S(\nu,d,n)$ is depicted as a contour plot in Fig.~\ref{fig:2}(b) for polynomial order $n=10$. Its minimum results in $\nu = 1.00\pm0.01$ ($\nu=1$ is the exact result~\cite{Aubry_80}) and $d = 0.54\pm0.01$ which is in excellent agreement with the numerical values obtained for static quasi-disordered systems~\cite{Li_Pixley_2016}. The error bars are estimated from a range of $\nu$ and $d$ with compatible values of $S$. 

While one would expect that the eigenvectors of the Floquet operator $U(T)$ and of the effective Floquet Hamiltonian to be equivalent when $T\ll1$ and $\lambda\ll1$ and, consequently, its scaling properties as well, what is remarkable is that sensitive quantities as the critical exponents do not suffer appreciable modifications when one is not exactly in this regime, namely at $T=0.5$ and $\lambda=1$.

\subsection{Wave packet propagation}\label{sec:IIB}
\begin{figure}[t]
\centering
\includegraphics[width=0.95\columnwidth]{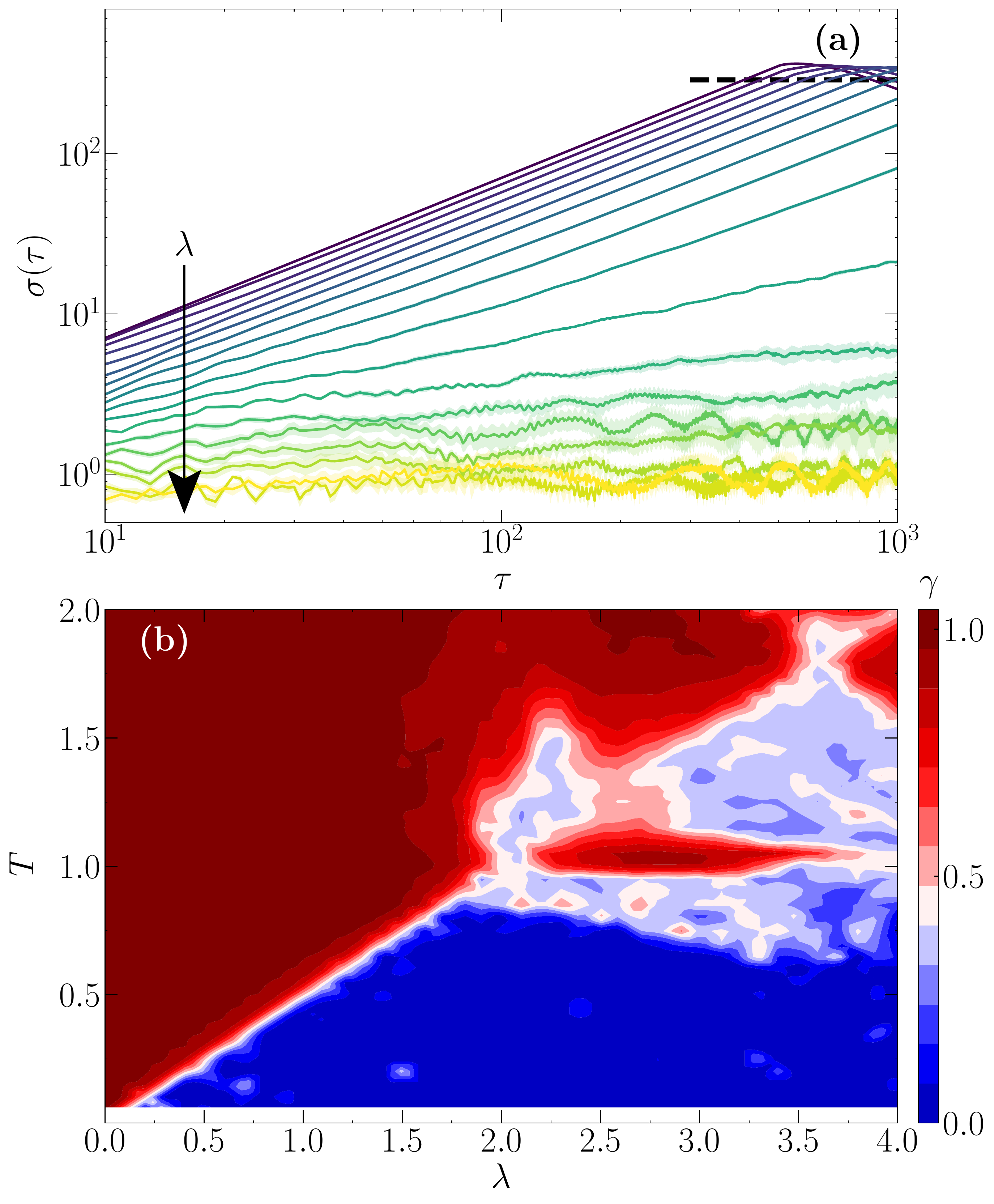}
\caption{(Color online) (a) The RMSD $\sigma$ as a function of the number of kicks $\tau$ for the AA model for various kick strengths $\lambda$ at fixed period $T = 0.5$ in a system with $L = 1000$. The values of $\lambda$ used are 0.0, 0.1, 0.2, \ldots, 1.5; 1.7 and 2.0 as schematically represented by the arrow, from darker to lighter colors. Each line is averaged over ten different realizations of $\varphi$, and the shaded area denotes the standard error. The lines can be fitted by an expression $\sigma \propto \tau^{\gamma}$, where $\gamma$ is the diffusion exponent. Black dashed line denotes the value of $\sigma$ for a plane wave. (b) Fitted $\gamma$ as a function of kick strength $\lambda$ and period $T$.}
\label{fig:3}
\end{figure}

Besides the properties of the Floquet eigenvectors, we also study the time-evolution of an initially fully localized state in the middle of the lattice: $|\psi_0\rangle = |L/2\rangle = \hat c^\dagger_{L/2}|\emptyset\rangle$. We follow the stroboscopic evolution by repeated application of the Floquet operator $\hat{U}(T)$, i.e., after one period $|\psi(T)\rangle = \hat{U}(T)|\psi_0\rangle$ and after a time $t=\tau T$, $|\psi(\tau T)\rangle = \hat{U}^{\tau}(T)|\psi_0\rangle$. To quantify the degree of the initial wave packet spreading, we use the root mean square of the displacement (RMSD), defined as 
\begin{eqnarray}
\sigma(\tau) = \left[\sum_i \left(i - L/2\right)^2 | \psi_i(\tau) |^2\right]^{1/2}. 
\label{eq:sigma}
\end{eqnarray}
An example of the time dependence of $\sigma(\tau)$ for the case of incommensurate kicks ($V_i = V_i^{\rm AA}$), at fixed period of the kicking $T = 0.5$ and for multiple kicking strengths $\lambda$, is presented in Fig.~\ref{fig:3}(a). Starting with the case of vanishing strength of the kick ($\lambda \to 0$), the wave packet propagates freely (ballistically) on the chain, resulting in a linear growth of the RMSD with time, i.e., $\sigma(\tau) \propto \tau$. Ballistic spreading is observed almost up to the critical value of the disorder strength $\lambda = 1$, where the dynamical Anderson localization transition occurs and the slope of $\sigma(\tau)$ drops to zero. In the localized regime ($\lambda > 1$), small oscillations of the $\sigma(\tau)$ occur on a time scale that is proportional to the diffusive spreading of the wave packet within the localization length~\cite{Haake}. The shaded area surrounding each curve depicts the standard error of the mean after the averaging with different phases $\varphi$. 

Due to the finiteness of the system, there is a maximum RMSD a particle can reach, given as $\sigma_{\rm max} = L/2$ (in case it is localized at either $i = 0$ or $L$ as well as a quantum superposition of both cases), whereas for our choice of initial state the wave function becomes delocalized, thus it oscillates around the value $\sigma_{\rm pw} = [(L^2 + 2)/12]^{1/2}$ obtained for a plane wave [dashed line in Fig.~\ref{fig:3}(a)]. By taking this limitation into account, we fit the exponent of the RMSD $\sigma \propto \tau^{\gamma}$ at intermediate time scales, avoiding the short transient behavior at early times and the maximum spreading at later ones that is manifest in finite-size systems. Values of $\gamma = 0, 1/2$ and $1$ indicate localization, diffusive and ballistic transport, respectively. The regime $0 < \gamma < 1/2$ ($1/2 < \gamma < 1$) corresponds to subdiffusion (superdiffusion). The phase diagram of the fitted $\gamma$ is shown in Fig.~\ref{fig:3}(b) and its similarity with the phase diagram extracted from IPR [Fig.~\ref{fig:1}(b)] is evident: the previous non-ergodic and ergodic regions in the $T$ vs. $\lambda$ diagram are connected to the localized and ballistic spreading of the wave packet. At large periods of the driving and large kick amplitudes, the effective Floquet time-independent Hamiltonian (Eq.~\ref{eq:floquet_hamil}), previously obtained no longer describes the stroboscopic time-evolution and a mixture of regimes of spreading ensues.

\section{Delocalization in non-periodically driven systems}
\label{sec:aperiodic}
Effects of decoherence in non-interacting systems that display dynamical localization, as in the paradigmatic quantum kicked-rotor, were also experimentally investigated in the past. Among these effects, experiments tried to induce additional spontaneous emission in the trapped atoms that emulate the quantum kicked rotor~\cite{Ammann_98}; others tried to induce noise in the amplitude of the periodic kicks~\cite{Klappauf_98}. In common, both mechanisms result in delocalization of the atomic wave function, leading to a quantum diffusive behavior. 

\begin{figure}[!t]
\centering
\includegraphics[width=0.95\columnwidth]{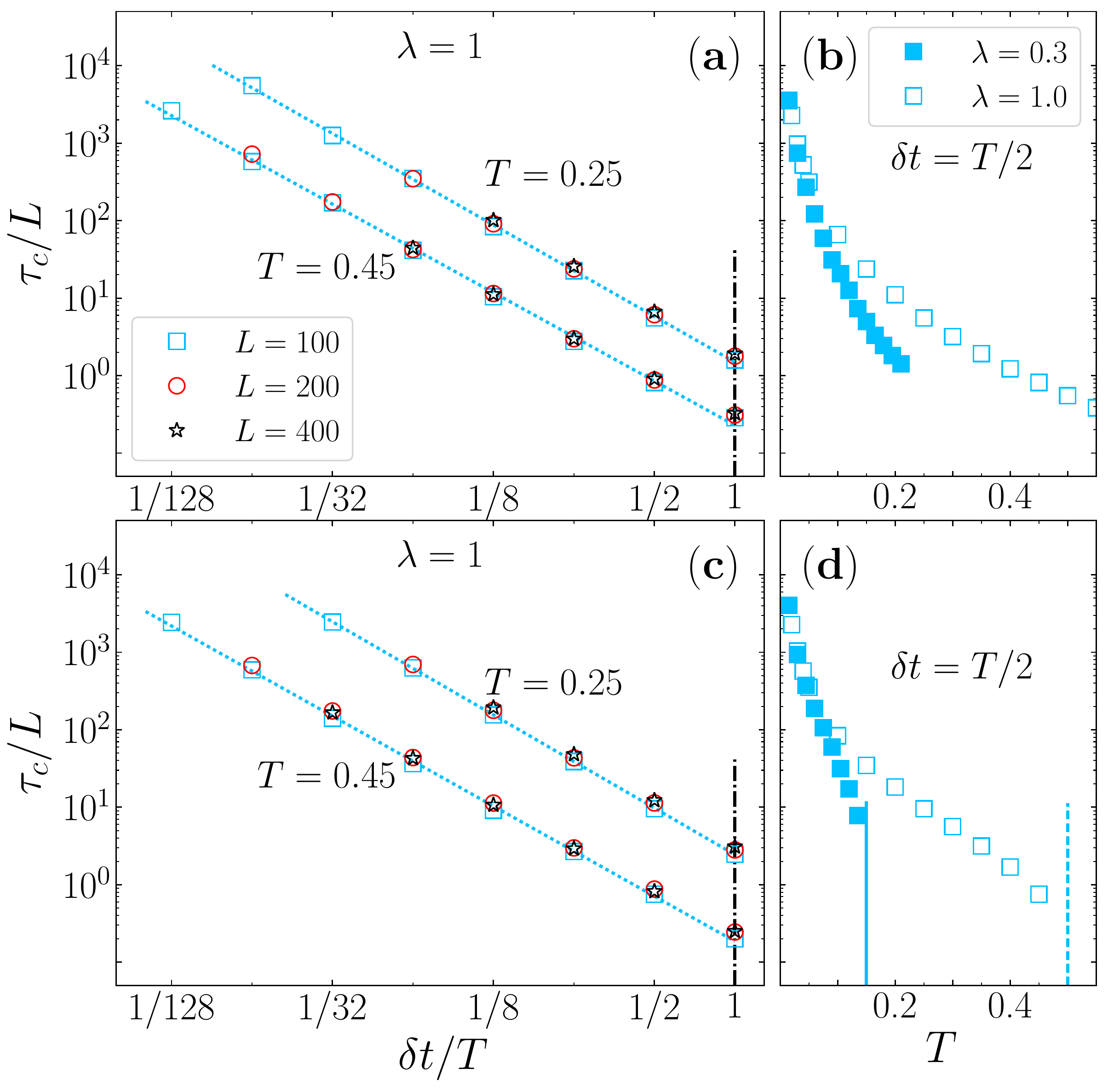}
\caption{(Color online) Characteristic number of kicks $\tau_c$ (normalized by the system size $L$) for the delocalization of eigenstates  due to aperiodicity for Anderson and Aubry-Andr\'e kicks in panels (a) and (c), respectively, as a function of the aperiodicity $\delta t/T$; note the log-log scale used. The extracted slopes of the linear fit (depicted by the dotted lines for the smallest system size) are $-1.85\pm0.05$ and $-1.93\pm0.02$ for $T = 0.45$ and $T = 0.25$, respectively, for Anderson kicks shown in panel (a) and $-1.91\pm0.02$ and $-1.99\pm0.01$ for $T = 0.45$ and $T = 0.25$, respectively, for Aubry-Andr\'e kicks shown in panel (c). Vertical dash-dotted lines denote the maximum amount of aperiodicity since $\delta t/T<1$. Panels (b) and (d) display the dependence of the ratio $\tau_c/L$ as a function of $T$ at fixed aperiodicity $\delta t = T/2$, for Anderson and Aubry-Andr\'e kicks, respectively. Vertical full (dashed) line depicts the transition value in the case of Aubry-Andr\'e kicks for $\lambda=0.3$ (1.0). The characteristic number of kicks for each point was determined from the average of 10, 100 and 1000 time and disorder realizations for the cases where $\tau_c > 1000$, $1000 > \tau_c > 100$ and $\tau_c < 100$, respectively.}
\label{fig:7}
\end{figure}

A third type of mechanism which was later shown to induce an unbounded growth of the total energy of the system is when there is noise in the time period of the kicks, observed in atomic and molecular quantum-kicked rotors~\cite{Oskay_2003, Bitter_2016, Bitter_2017}. Analogously, here we consider a similar situation, by allowing that the perfect periodicity of the kicks suffers from deviations in our kicked lattice model. To model that, we assume that the time between two consecutive kicks $T_{\tau}$ is a stochastic variable distributed with equal probability between times $T - \delta t$ and $T + \delta t$. Thus, the time of the ${\tau}$-th kick is given as $t_{\tau} = t_{{\tau}-1} + T  + \delta t_{\tau}$\footnote{An alternative scheme where $t_{\tau} = \tau T  + \delta t_{\tau}$, with $\delta t_{\tau} \in (-\delta t, \delta t)$ and $\delta t < T/2$, gives qualitatively the same results as the scheme considered in the text. Note that in this alternative scheme, the time between two consecutive kicks is given as $T_{\tau} = T + \delta \tilde{t}_{\tau}$, where $\delta \tilde{t}_{\tau} = \delta t_{\tau-1} - \delta t_{\tau}$, i.e., $\delta \tilde{t}_{\tau}$ is a convolution of two stochastic distributions. For example, the extracted power law exponents $\alpha$ in the expression $(\tau_c/L)\propto (\delta t/T)^\alpha$, as given in Fig.~\ref{fig:7} (c) are $-1.88\pm0.01$ and $-1.98\pm0.01$ for $T = 0.45$ and $T = 0.25$, respectively, for Aubry-Andr\'e kicks which is in close agreement with the results for the other aperiodic scheme used.}, with $\delta t_{\tau} \in (-\delta t, \delta t)$, and $\delta t < T$ to obey causality. This timing noise scheme was used in the experimental study of a quantum kicked rotor~\cite{Oskay_2003}. The time evolution operator after ${\tau}$ kicks is 
\begin{eqnarray}
 \hat{U}_{\tau} = \hat{U}(T_{\tau}) \hat{U}(T_{{\tau}-1}) \cdots \hat{U}(T_1),
 \label{eq:U_tau}
\end{eqnarray}
with $T_{\tau} = t_{\tau} - t_{{\tau}-1}$ and $\hat{U}(T_{\tau}) = e^{-{\rm i} \hat{H}_0 T_{\tau}} e^{-{\rm i} \lambda \hat{V}}$. Note that the randomness in the propagation time $T_{\tau}$ can be similarly interpreted as randomness in the hopping energy $J$. This would correspond to a situation where the energy scale of the problem changes from kick to kick in a stochastic manner, affecting the tunneling probabilities of each bond in the lattice.

\begin{figure}[t]
\centering
\includegraphics[width=0.95\columnwidth]{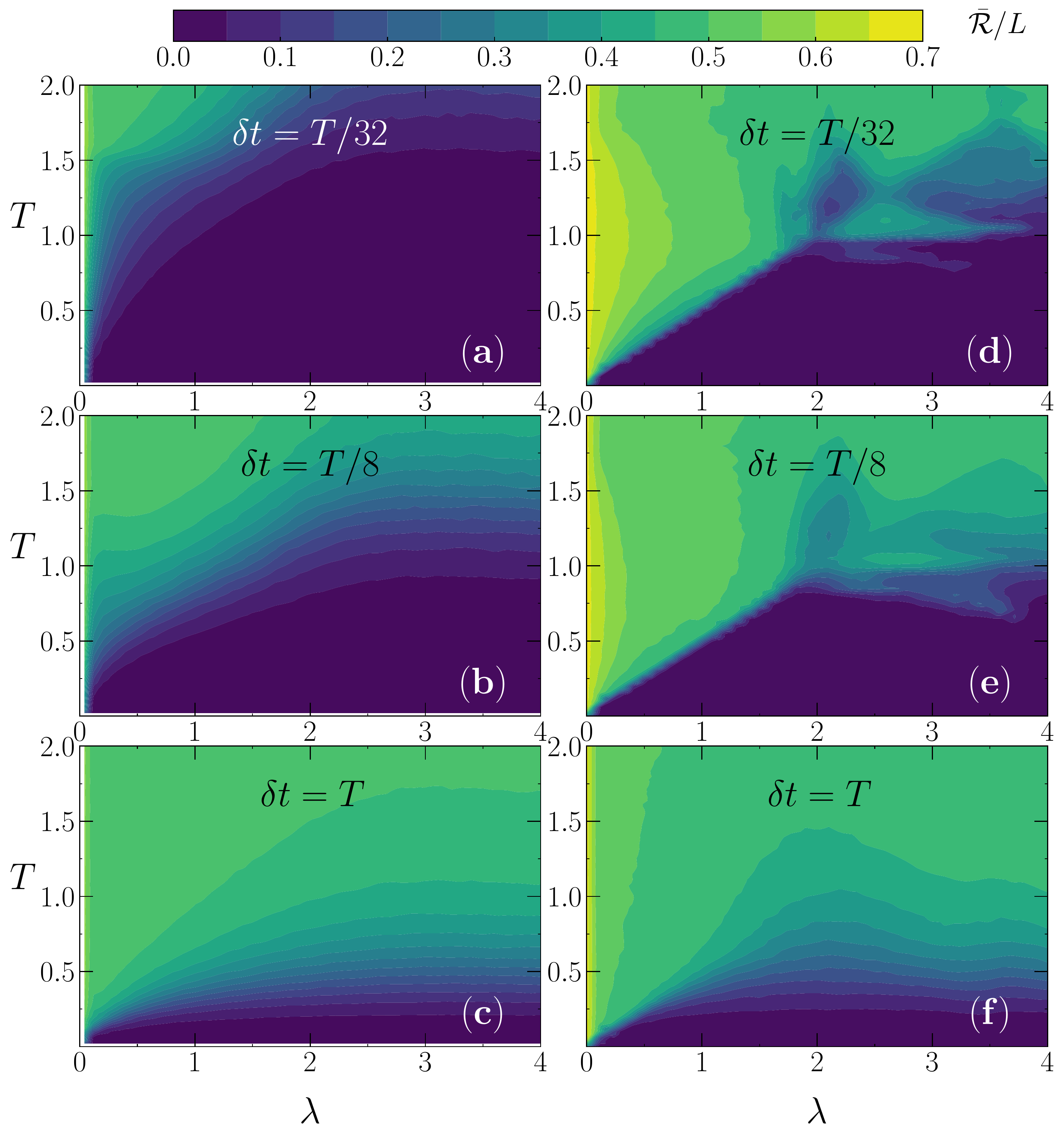}
\caption{(Color online) Average IPR as a function of the $\lambda$ and $T$ after 1000 random kicks in the quantum-kicked Anderson and Aubry-Andr\'e model are shown. We compare three different aperiodicity magnitudes $\delta t = T/32, T/8$, and  $T$ in panels (a), (b) and (c) for Anderson type kicks and in panels (d), (e) and (f) for Aubry-Andr\'e kicks, respectively. The system size used is $L = 1000$ and we average over 10 realizations of both disorder and time sequence.}
\label{fig:4}
\end{figure}

In the following, we study the average IPR after a number $\tau$ of aperiodic kicks, obtained from exact diagonalization of Eq.~\eqref{eq:U_tau}. We start by checking how an increasing aperiodicity $\delta\tau$ affects the ``phase diagram'' [Fig.~\ref{fig:1}] originally obtained for the periodic driving. This is reported in Fig.~\ref{fig:4} for a fixed number ($\tau = 1000$) of aperiodic kicks. We see that at fixed small aperiodicity [$\delta\tau = T/32$ in panels (a) and (d)] the $T$ vs. $\lambda$ diagram is mostly affected in the high average period regime in comparison to the periodic case, where the average IPR increases for both types of kicks considered. With larger aperiodicity, the region of enhanced IPR progresses towards lower average periods $T$, signaling that the dynamical localization becomes less robust when $\delta\tau$ grows.  

\begin{figure}[t]
\centering
\includegraphics[width=0.95\columnwidth]{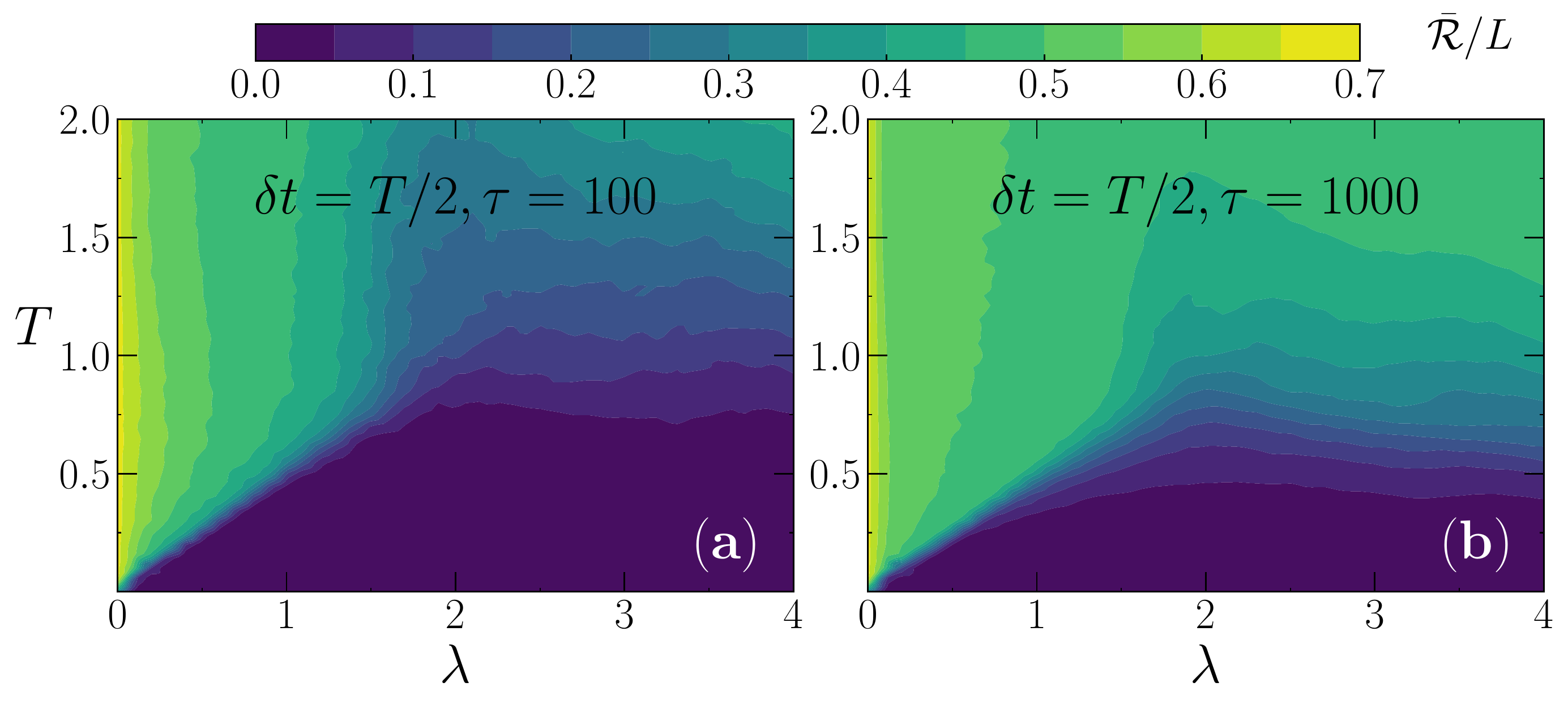}
\caption{(Color online) Average IPR as a function of $\lambda$ and $T$ after $\tau = 100$ and $\tau = 1000$ aperiodic kicks in the quantum-kicked Aubry-Andr\'e model is shown in panels (a) and (b), respectively. The system size used is $L = 1000$, aperiodicity $\delta t = T/2$, and we average over ten realizations of both disorder and time sequence.}
\label{fig:5}
\end{figure}

Now, we focus on another aspect, by fixing the aperiodicity at $\delta\tau=T/2$ and comparing the IPR phase diagram for different number of kicks, as shown in Fig.~\ref{fig:5}, for the case of AA potentials. We note that the larger number of aperiodic kicks increases the area of enhanced average IPR. Thus, both the aperiodicity and a larger number of kicks suggest to lead the system to disrupt its original dynamical localization.

\begin{figure}[t]
\centering
\includegraphics[width=0.99\columnwidth]{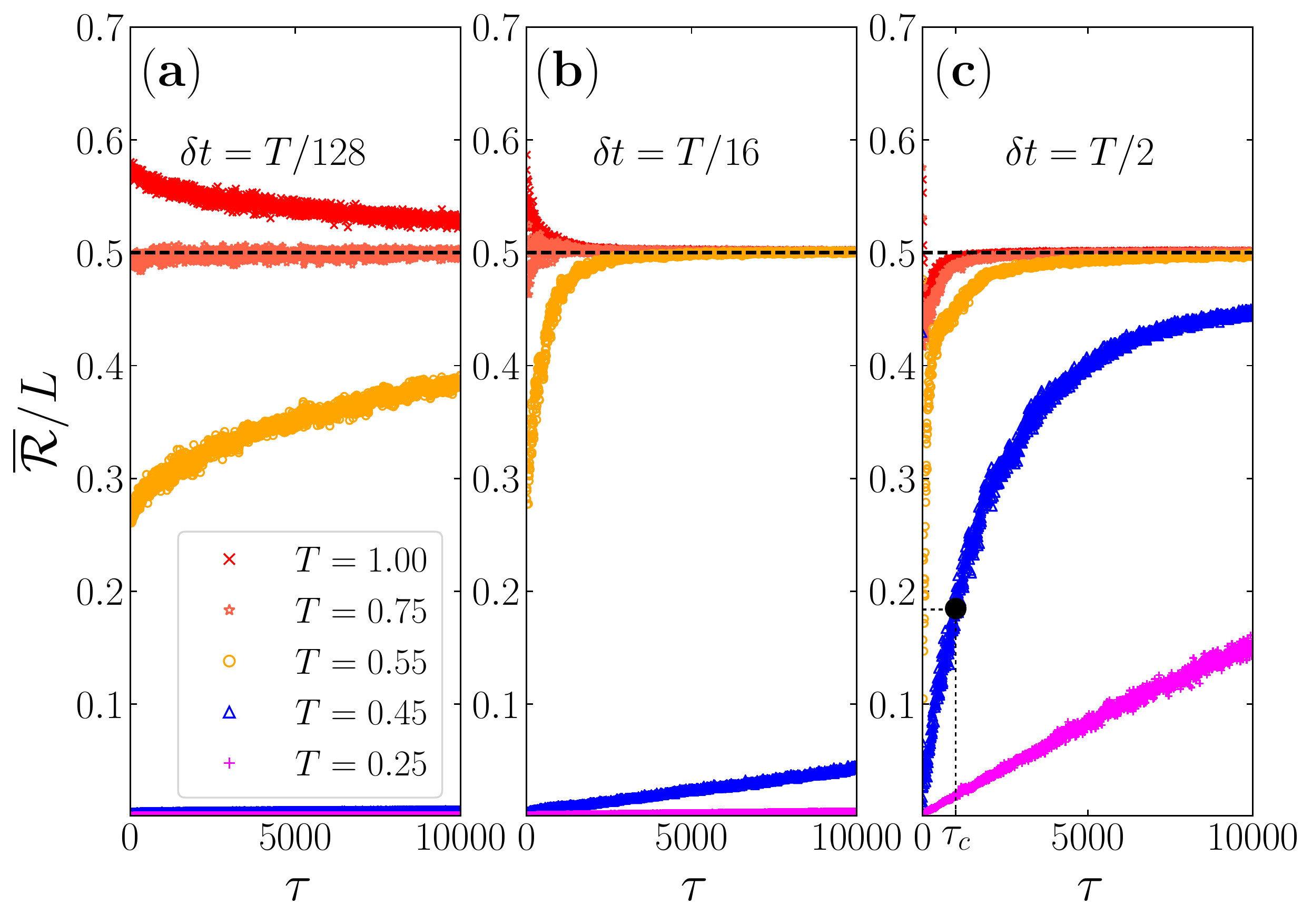}
\caption{(Color online) Average IPR as a function of the number of kicks in the quantum-kicked Aubry-Andr\'e model with $\lambda = 1$ in a lattice with $L = 1000$. We compare three different aperiodicity magnitudes $\delta t = T/128, T/16$, and  $T/2$ in panels (a), (b) and (c), respectively. The (black) dashed horizontal line represents the average IPR of a fully delocalized wave function obtained from a random matrix belonging to a GOE. In panel (c), we exemplify the characteristic number of kicks $\tau_c$ in the delocalization process for the mean periodicity $T = 0.45$ (see text).}
\label{fig:6}
\end{figure}

Both analyses provide a good qualitative picture of the localization (and its progressive destruction) under the aperiodic driving. However, a more quantitative study is required to see whether a characteristic number of kicks that leads to delocalization can be obtained. For that, we study the evolution of the average IPR with the number of kicks $\tau$ in Fig.~\ref{fig:6}, for increasing magnitudes of the aperiodicity $\delta\tau$. We fix the kick strength $\lambda$ to 1, where the corresponding critical period separating the localized and delocalized behavior if the kicks were periodic in time is equal to $T_c = 0.5$. In Fig.~\ref{fig:6}(a), we start with a very small perturbation of the periodic case $\delta t = T/128$ and note that the IPR begins to converge to 0.5 as the number of kicks is increased. This convergence occurs with smaller number of kicks when increasing the aperiodicity, as can be seen in Figs.~\ref{fig:6}(b) and \ref{fig:6}(c). This limit result could be interpreted as being related to a fully delocalized state in the (real-space) basis: If one starts with a random matrix $A$ (whose eigenstates by definition are maximally delocalized) belonging to the Gaussian orthogonal ensemble (GOE), we notice that the average mean IPR of the matrix $e^{{\rm i}A}$ is exactly 0.5. As a consequence, delocalization sets in once the number of kicks is large enough. However, for periods smaller than $T_c$ we notice that the growth of the average IPR towards the 0.5 limit is rather slow for small values of $\delta t$. Yet, if one increases the aperiodic effects [Figs.~\ref{fig:6}(b) and ~\ref{fig:6}(c)] the growth rate deep in the once localized regime increases, and the dynamical localization is progressively erased. This absence of localization in real space due to phase decoherence of the eigenvectors is similar to what is observed in the atomic and molecular quantum-kicked rotors~\cite{Oskay_2003, Bitter_2016, Bitter_2017}, where the original dynamically localized state in momentum or in angular momentum is destroyed after the noise in time is introduced.

Although the increase of the IPR towards the delocalized result (0.5) after sufficient number of aperiodic kicks is suggestive of overall delocalization, one needs to check the sensitivity of the results to finite-size effects, and to see whether there exists a finite disorder in time that triggers the breakdown of dynamical localization. To test both conditions, we define a characteristic number of kicks $\tau_c$, at which the average IPR reaches $1/e$ of the final value, {\it i.e.} $\overline{\cal R}(\tau_c)/L = 0.5/e$\footnote{Other definitions of characteristic critical values for models of exponential saturation could be used as, for example,  $\overline{\cal R}(\tilde{\tau}_c)/L  = 0.5(1-e^{-1})$. We notice though that this latter definition results in computationally prohibitively large values of the characteristic number of kicks related to delocalization, if one investigates the situation that were originally deep in the localized regime for periodic kicks.}. In Figs.~\ref{fig:7}~(a) and \ref{fig:7}~(c), we show the dependence of $\tau_c$ on different kick aperiodicities $\delta t$ for different system sizes and mean kick periods $T$, in the cases of Anderson and Aubry-Andr\'e kicks, respectively. We notice that the characteristic number of kicks $\tau_c$ is nearly proportional to the system size $L$ in both schemes, since $\tau_c/L$ does not substantially change for different values of $T$ and $\delta t$. Besides, the ratio $\tau_c/L$ displays a power law behavior on the aperiodicity, $(\tau_c/L)\propto (\delta t/T)^\alpha$, with $\alpha\approx-2$, as shown by the dashed lines that depict the fitting using this functional form. This tells us that a smaller number of kicks are necessary to induce delocalization once the disorder in the time periodicity is enhanced.

\begin{figure*}[!tbh]
\centering
\includegraphics[width=0.8\textwidth]{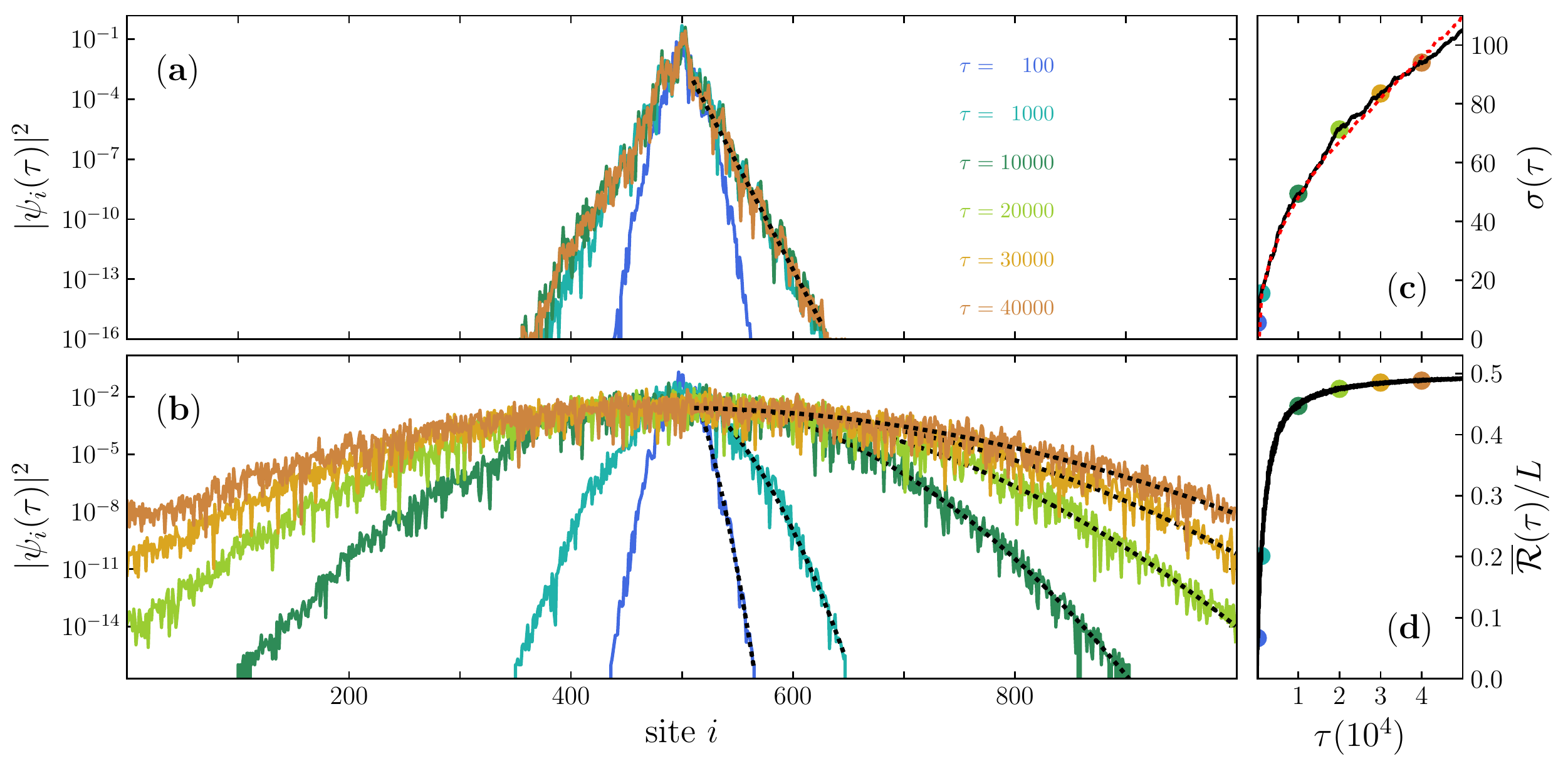}
\caption{(Color online) Probability density distributions of an initially localized state after various numbers of kicks, for periodic and aperiodic case, with $\delta t = T/2$, are shown in panels (a) and (b), respectively. The probabilities are displayed after $\tau = 100, 1000, 10000, 40000$ kicks in panel (a) and after $\tau = 100, 1000, 10000, 20000, 30000, 40000$ aperiodic kicks in panel (b); note the logarithmic scale on the y-axis. Dotted lines in panels (a) and (b) denote exponential and Gaussian fits, respectively. Panel (c) shows the wave function spreading $\sigma$ as a function of the number of aperiodic kicks, calculated by Eq.~\ref{eq:sigma} in full black line and obtained from fitting a Gaussian function to the probabilities (red dotted line). Panel (d) shows the average IPR as a function of the number of aperiodic kicks. The characteristic number of kicks in this case is $\tau_c \approx 756$. The circles in panels (c) and (d) highlight the values of $\sigma$ and $\overline{\cal R}/L$, respectively after the same number of kicks as in panel (b). The parameters used are $\lambda = 1$, $T = 0.45$ and system size $L = 1000$.}
\label{fig:8}
\end{figure*}

Furthermore, we investigate how the delocalization is affected at fixed aperiodicity $\delta t$. If one starts deeper in the formerly dynamically localized regime, were the kicks periodic in time [i.e., for values of the ratio $\lambda/T$ much smaller than the critical value for the Aubry-Andr\'e kicks, $(\lambda/T)_c = 2$] a larger amount of kicks is necessary to induce delocalization. This is exemplified by the two sets of points in Fig.~\ref{fig:7}~(c), for $T=0.25$ and $0.45$. The critical period at the fixed kick strength $\lambda = 1$ is $T_c = 0.5$, and the number of kicks to reach $\tau_c$ is approximately one order of magnitude smaller in the latter. We systematically investigate this in Fig.~\ref{fig:7}~(d), where we show the ratio $\tau_c/L$ as a function of $T$ at fixed $\delta t =T/2$. This ratio decreases monotonically when approaching the dynamical localization transition marked by the vertical full (dashed) line for $\lambda = 0.3$ (1.0).

Similarly, in the case of Anderson-type kicks [Figs.~\ref{fig:7}~(a) and \ref{fig:7}~(b)], we see that a larger $\tau_c$ is necessary to induce delocalization if one is close to the robust limit of dynamical localization for periodic kicks, namely, at high frequencies ($T\to0$). Overall, these results suggest that any nonzero aperiodicity $\delta t$ leads to eventual delocalization after sufficient number of aperiodic kicks. However, even for large aperiodicity, we see that the number of kicks necessary to destroy localization rises considerably when both moving further in the localized regime (with smaller periodicity $T$) and when increasing the kicking strength (with larger $\lambda$).

Finally, we study the wave packet propagation that is manifestly relevant for experiments. As in Sec.~\ref{sec:IIB}, we set an initial localized state in the middle of the chain and follow the stroboscopic evolution after applying a series of aperiodic kicks. In Figs.~\ref{fig:8}(a) and \ref{fig:8}(b), we show snapshots of the probability distributions after different number of kicks $\tau$, for a periodic and a non-periodically kicked system, respectively. In the periodic case, the localized state initially spreads until it becomes exponentially localized, when the wave packet reaches the localization length for the set of parameters $T$ and $\lambda$ used (0.45 and 1.0, respectively), after a sufficient number of kicks. One can use an exponential fitting of the form $|\psi_i(\tau)|^2 \propto \exp{(|i-L/2|/\xi_{\rm loc})}$ to find the localization length to be $\xi_{\rm loc} = 4.2 \pm\ 0.1$, which is much smaller than the actual system size, $L=1000$.

In contrast, the aperiodic kicking causes diffusion of the initial state, and the probability distribution can be approximated by a Gaussian function of the form $|\psi_i(\tau)|^2 = [2 \pi \sigma^2(\tau)]^{-1/2} \exp[-\frac{(i-L/2)^2}{2 \sigma^2(\tau)}]$ (dotted lines) at subsequent times. The width $\sigma$ of the wave packet has a power-law dependence on the number of kicks in the case of diffusive spreading. This is confirmed in Fig.~\ref{fig:8} (c), where we compare the fitted wave function spreading $\sigma_f(\tau)$ with the one calculated by Eq.~\ref{eq:sigma}. Both curves give the dependence $\sigma(\tau) \propto \tau^{\gamma}$ with $\gamma = 0.49$ and 0.5 for calculated and fitted data, respectively, confirming the diffusive aspects in the non-periodic setting of the kicks. Note that for a Gaussian wave packet one can also calculate the IPR ${\cal R}_G$ in the limit $\sigma(\tau) \ll L$, which gives ${\cal R}_G(\tau) = 2 \sqrt{\pi} \sigma(\tau)$, thus the IPR of the Gaussian state is proportional to $\sigma(\tau)$ as presented in Fig.~\ref{fig:8} (c). Lastly, we revisit in Fig.~\ref{fig:8}(d) the dependence on the number of kicks of the average mean IPR for the aperiodic case. This is to be compared to Fig.~\ref{fig:6}(c), with a similar set of parameters. We now identify, however, the increasing IPR of the eigenstates of the time evolution with the correspondent density distributions by the circles with compatible colors of Fig.~\ref{fig:8}(b), representing the same number of kicks. We see that after the wave packets start reaching the ends of the lattice, the saturation to a fully delocalized state ($\overline {\cal R}/L\sim0.5$) is asymptotically approached. In this regime, this would correspond to the flattening of the Gaussian distribution which will be only obtained when $\tau\to\infty$.

\section{Summary}
\label{sec:summary}
We studied the phenomenon of dynamical localization and its breakdown in a driven lattice model. The driving, consisted of kicks in time that quench the onsite energies of the lattice, leads to localization in real space provided that it is periodic and its frequency is large. We evaluate this for two types of kicks, where the onsite energies are either instantaneously disordered or when they are quasi-periodic in space.  By finding a time-independent effective Hamiltonian in the high-frequency regime, we identify the transitions to dynamical localization to be correspondent to the ones in static cases, provided one rescales the kick amplitudes by the period of the driving. When we consider the kicks to be no longer regularly spaced in time, we notice an unbounded increase of the widths of the probability distributions of the wave packets in real space after successive kicks, when starting from an initially localized state. That is similar to the observed decoherence effects obtained for atomic and molecular quantum kicked rotors. Moreover, we notice that an effective delocalization takes more kicks to develop if the system is in regimes of parameters that are deep in the dynamically localized phase, were the kicks periodic in time. Although the number of kicks that leads to delocalization increases, our results suggest that any finite aperiodicity lead to the destruction of localization for long enough drivings and this conclusion is not systematically affected by finite size effects.

\begin{acknowledgments}
The authors acknowledge discussions with R.~Fazio, G.~Santoro and C.~Cheng. TC is supported by the National Natural Science Foundation of China (NSFC) Grant No. 11650110443; RM is supported by NSFC Grants Nos. U1530401, 11674021 and 11650110441 and PDS acknowledges partial support from FCT through grant UID/CTM/04540/2013. The computations were performed in the Tianhe-2JK at the Beijing Computational Science Research Center (CSRC).

\end{acknowledgments}

\bibliography{kicks}

\end{document}